\documentclass{aa}     
\usepackage[dvips]{graphics}
\usepackage{latexsym}
\begin{document}

\newcommand{\zs} {$\zeta_\star$}
\newcommand{\vsini} {$v\,\sin i$}
\newcommand{\Sc} {Sect.}
\newcommand{\cmc} {cm$^{-3}$}
\newcommand{\cmq} {cm$^{-2}$}
\newcommand{\csqg} {cm$^2$ g$^{-1}$}
\newcommand{\kms} {km s$^{-1}$}
\newcommand{\myr} {$M_\odot$ yr$^{-1}$}
\newcommand{\Myr} {$M_\odot$ yr$^{-1}$}
\newcommand{\um} {$\mu$m}
\newcommand{\mic} {$\mu$m}
\newcommand{\sbr} { erg cm$^{-2}$ s$^{-1}$ sr$^{-1}$}
\newcommand{\sbu} { erg cm$^{-2}$ s$^{-1}$ sr$^{-1}$}
\newcommand{\Lsun} {L$_\odot$}
\newcommand{\Msun} {M$_\odot$}
\newcommand{\MJ} {M$_{\rm J}$}
\newcommand{\Teff} {T$_{\rm{eff}}$}
\newcommand{\Tstar} {T$_{\rm{eff}}$}
\newcommand{\Lstar} {L$_\star$}
\newcommand{\Rstar} {R$_\star$}
\newcommand{\Mstar} {M$_\star$}
\newcommand{\Md} {M$_{\rm D}$}
\newcommand{\Rd} {R$_{\rm D}$}
\newcommand{\Rin} {R$_i$}
\newcommand{\pms} {pre-main--sequence}
\newcommand{\DV} {$\Delta V$}
\newcommand{\Dm} {$\Delta m$}
\newcommand{{\AV}} {A$_{\rm V}$}
\newcommand{{\AJ}} {A$_{\rm J}$}
\newcommand{{\1}} {Cha H$\alpha$1}
\newcommand{\rhooph} {$\rho~$Oph}
\newcommand{\mir}{F$_{14.3}$/F$_{6.7}$}

\newcommand{\simless}{\mathbin{\lower 3pt\hbox
      {$\rlap{\raise 5pt\hbox{$\char'074$}}\mathchar"7218$}}} 
\newcommand{\simgreat}{\mathbin{\lower 3pt\hbox
     {$\rlap{\raise 5pt\hbox{$\char'076$}}\mathchar"7218$}}} 

\title{Exploring Brown Dwarf Disks in $\rho$ Oph
\thanks{Partly based on observations collected at the Italian Telescopio
Nazionale Galileo (TNG) operated on the island of La Palma by the Centro
Galileo Galilei of INAF (Istituto Nazionale di 
Astrofisica) at the Spanish Observatorio del Roque de los Muchachos of the
Instituto de Astrofisica de Canarias, and at the European Southern Observatory,
La Silla and Paranal, Chile.}}

\author {Antonella Natta\inst{1}, Leonardo Testi\inst{1}, Fernando Comer\'on\inst{2},
Ernesto Oliva\inst{1,3}, Francesca D'Antona\inst{4}, Carlo Baffa\inst{1}, Giovanni
Comoretto\inst{1}, Sandro Gennari\inst{1}}
\institute{
    Osservatorio Astrofisico di Arcetri, INAF, Largo E.Fermi 5,
    I-50125 Firenze, Italy
\and
ESO, Karl-Schwarzschild-Strasse 2, D-85748 Garching Bei M\"unchen, Germany
\and
TNG and Centro Galileo Galilei, INAF, P.O. Box 565, E-38700, Santa Cruz de La Palma, Spain
\and
Osservatorio Astronomico di Roma, INAF, via Frascati 33, I-00044 Roma, Italy
}

\offprints{natta@arcetri.astro.it}
\date{Received ...; accepted ...}
 
\titlerunning{BD disks}
\authorrunning {Natta et al.}

\begin{abstract}
{This paper discusses  evidence for and properties of disks associated
to brown dwarfs in the star-forming region \rhooph.  We selected nine objects from the ISOCAM survey of Bontemps et al.~\cite{Bonea01}
that have detections in the two mid-infrared
bands (6.7 and 14.3 \um), relatively low extinction and low luminosity. 
We present  low-resolution near-infrared spectra in the J, H and K bands, 
and determine for each source spectral type, extinction, effective temperature and luminosity by comparing the spectra to those of field dwarfs and
to the most recent model stellar atmospheres. The results indicate
that eight objects have spectral types M6--M7.5, effective temperature
of 2600--2700 K, one has a later spectral type (M8.5) and lower temperature
(about 2400 K). The derived extinctions range between \AV$\sim$2 and 8 mag.  
The location of the objects on the HR diagram, in spite of the uncertainties of the evolutionary tracks for young objects of substellar mass, 
indicates that all the objects are very young and have masses below about 
0.08 \Msun. The coolest object in our sample  has mass in the range 8-12 \MJ\
(0.008--0.012 \Msun).
In all cases, the mid-infrared excess is consistent with the predictions of models of disks irradiated by the central object, showing
 that circumstellar disks are commonly associated to young brown dwarfs and planetary-mass objects.
Finally,
we discuss possible variations of the disk geometry among different objects,
as well as the possibility of using these data to discriminate between various formation scenarios. 
}
\end{abstract}

\maketitle

\section{Introduction}

A large number  of objects with sub-stellar mass  are now known,
with masses ranging  from the hydrogen burning limit that
divides stars from brown dwarfs (BDs; \Mstar$\simless 
0.075$ \Msun)  to values comparable to the mass of giant
planets and below the deuterium burning limit 
($\simless$ 0.013 \Msun).
Their discovery 
in regions of star formation
has provoked  an intense debate on the formation mechanism
of such objects. Do they form, as solar mass stars do, from the
collapse of a molecular core (Shu et al.~\cite{Shu87})? 
Are they stellar embryos, 
whose further growth is prevented by dynamical
ejections from small stellar systems (Reipurth \& Clarke~\cite{RC01};
Bate et al.~\cite{Bate02})?
 Or are they ``planets", i.e., objects that form
in gravitationally unstable regions of circumstellar disks
(Papaloizou \& Terquem~\cite{PT01};  Lin et al.~\cite{Linea98})?
Is there a single formation process for all substellar objects?
What is the lowest mass for the gravitational collapse mechanism?

A crucial contribution to this debate is expected from studies  of the circumstellar disks (if any) associated with sub-stellar objects, since different theories make very different predictions.   Disks are a necessary
step in any formation mechanism that involves
accretion from a parental core. If BDs  form from core
collapse, they should be associated to disks similar in properties
to those found around low mass \pms\ stars (T Tauri stars; TTS).
A prediction of the stellar embryo theory is that the disks should
be truncated by the ejection mechanism, so that  they should be small and short-lived. 
In the ``planetary" hypothesis,  any circumstellar disk should be
even less substantial.

In some young BDs, emission in excess
of that due to the photosphere has been detected in the near
(Oasa et al.~\cite{OTS99}; Muench et al.~\cite{MLAL01}; Wilking
et al.~\cite{WGM99}) and  mid-infrared (Persi
et al.~\cite{Pea00}; Comer\'on et al.~\cite{Comea98}, \cite{CNK00}; 
Bontemps et al.~\cite{Bonea01}), and has been interpreted, by
analogy with TTS,  as evidence
for circumstellar disks.
In an earlier study (Natta \& Testi \cite{NT01}; Paper I),
we discussed the properties of three objects in Chamaeleon I
for which we could find in the literature ground-based spectroscopy and
photometry as well as ISOCAM measurements at 6.7 and 14.3 \um\
(Comer\'on et al.~\cite{CNK00}; Persi et al.~\cite{Pea00}). One of these objects is
a bona-fide BD,  while
the two others are close to the threshold between stars and BDs.
We found that the excess emission was clearly
detectable only in the mid-infrared, because the stellar photosphere
overwhelms the disk emission in  the three near-infrared bands.
The observed SEDs are well described by disk models similar to
those of TTS,  assuming that
the heating is due to irradiation from the central star.

This first result provides strong support for the idea that BDs form
like stars, from the contraction of a molecular core. 
Hence, we decided to extend our study of disk properties
to a larger number
of substellar mass objects in regions of star formation,
possibly down to objects of few Jupiter masses.
With this in mind, we  selected a small but well defined sample
of nine objects in the \rhooph\ region, that were detected by ISOCAM
at both 6.7 and 14.3 \um\ (Bontemps et al.~\cite{Bonea01}). We 
obtained near-infrared spectra for all of them (see \S 2),
which we used to derive the basic parameters of the central objects,
namely effective temperature, luminosity and mass (\S 3). Because of the
adopted selection criteria, all of these objects have   excess emission
in the mid-IR. We model the expected disk emission
for each object and show the results in \S 4.
We discuss the implications of our findings in \S 5 and present
conclusions in \S 6.

\section {Observations and data reduction}

\subsection {Selection criteria}

We chose nine BD candidates from the sample of 
Class~II objects detected at both 6.7 and 14.3 \um\ by Bontemps et
al.~(\cite{Bonea01}). We selected all  objects with visual extinction
less than $\sim$8.5~mag and luminosity less than $\sim$0.04~\Lsun\
according to Bontemps et al.~(\cite{Bonea01}). The first criterion 
ensures the possibility of obtaining high signal to noise specta
across the entire near infrared range. The low luminosity was required
to increase the chance of selecting objects in the range of masses we are 
interested in.

\begin{figure}
\resizebox{\hsize}{!}{\includegraphics{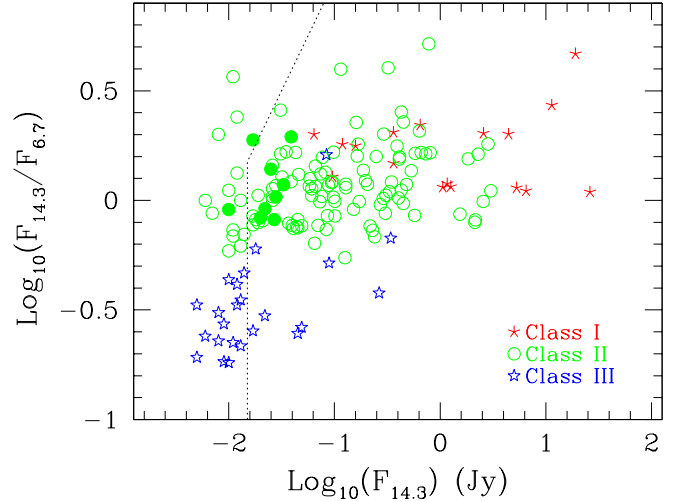}}
\caption{ISOCAM color-magnitude plot (adapted
from Bontemps et al.~\cite{Bonea01}). 
The symbols are
 asteriscs for  Class I sources, open circles for Class II sources, 
 stars for Class III sources. Filled circles show the objects in our sample.
The dotted line indicates the ISOCAM completeness limit.
 }
\label{fflcol}
\end{figure}

The location of the selected objects in the ISOCAM color-magnitude
diagram is shown in Figure~\ref{fflcol} (filled circles).
We note that all the objects are close to or below the completeness limit of
the ISOCAM survey, as expected for such low luminosities. 
In colors,
our sample
span the whole range covered by the Class~II objects (essentially
classical TTS).

Some of the selected sources were  known from previous studies to be 
very low-mass objects. In Table~1 we give the ISO source number,
the J2000 coordinates, other designations and references to previous studies.
Finding charts are provided in Appendix~\ref{images}.
We will comment on the comparison between the literature source parameters
and those derived in this paper in Section~\ref{sdiscspar}.
The results on one of the sources in our sample (\#033) have been already
presented in Testi et al.~(\cite{Tea02a}, hereafter Paper~II);
they have been re-analyzed and reported 
again here for an easier comparison with the rest of the sample.

\subsection {Near-infrared spectroscopy}

Near-infrared spectra for the objects in our sample 
were acquired in the period July 4--9, 2001 at the Telescopio Nazionale
Galileo (TNG), using the multi-mode Near-Infrared Camera Spectrograph (NICS;
Baffa et al.~\cite{Bea01}). The Amici device (Oliva~\cite{O00}),
a prism based, high-throughput
optical element unique to NICS, was used as disperser, coupled with a
0.5\arcsec\ wide, 4.2\arcmin\ long slit;
the resulting effective resolution is approximately
$\Delta\lambda/\lambda\sim$100, approximately constant across the entire
spectral range (0.85--2.45 \um). An identical instrumental configuration was used for the 
observations of field dwarfs of known spectral type (Testi et al.~\cite{Tea01};
\cite{Tea02b}). Data reduction and calibration was performed as described in
Testi et al.~(\cite{Tea01};~\cite{Tea02a}).

\subsection {Broad-band photometry}

On August 1 and 3 2001, we obtained moderately deep Gunn-i integrations 
using DFOSC and the Danish 1.54~m telescope at the ESO La~Silla
observatory. Following standard bias, flat fielding and 
sky subtraction, typically 6 individual 15~min dithered frames
were coadded to produce the final images. Photometric calibration
was achieved by observing a set of stars from the Landolt~(\cite{L92})
catalogue, for which i-band AB magnitudes were computed using the 
transformations given in Fukugita et al.~(\cite{Fea96}).
Given the uncertainties in the transformations and the non perfect
observing conditions, the uncertainties in the  photometry are
rather large.

For all sources near-IR J,H,K$_s$ photometry is available from the 2MASS
second incremental data release. Additional L' and R-band photometry 
were taken from Comer\'on et al.~(\cite{Comea98}).

\section {Spectral classification and stellar properties}

\begin{table}
\begin{flushleft}
\caption{ Sample objects and i-band photometry}
\vskip 0.1cm
\begin{tabular}{ccccc}
\hline\hline
(1)   & (2)    & (3)& (4)  & (5)      \\
Object&\multicolumn{2}{c} {Coordinates}&  i-band  & Other \\
(ISO\#)& \multicolumn{2}{c}{(J2000.0)}  &  (AB mag)&   Names \\
\hline
     &            &          &                 &         \\
023  & 16 26 18.8 & -24 26 09& 20.34 $\pm$0.15 & SKS1-10 \\
030  & 16 26 21.4 & -24 25 59& 16.50 $\pm$0.10 & SKS1-13\\
     &            &          &                 &GY5 \\
032  & 16 26 21.7 & -24 44 43& 16.26 $\pm$0.10 & --     \\
033  & 16 26 22.2 & -24 24 05& 21.73 $\pm$0.20 & SKS3-13\\
     &            &          &                 & GY11\\
102  & 16 27 06.5 & -24 41 50& 15.75 $\pm$0.10 & GY204\\
160  & 16 27 37.4 & -24 17 58& --              & --    \\
164  & 16 27 38.6 & -24 38 39& 18.18 $\pm$0.10 & SKS1-49\\
     &            &          &                 &   GY310\\
176  & 16 27 46.3 & -24 31 41& --              & GY350\\
\smallskip
193  & 16 28 12.2 & -24 11 37& --              & --       \\
\hline
\hline
\label{logobs}
\end{tabular}
\end{flushleft}
References for Column 5. SKS: Strom et al.~(\cite{SKS95});
GY: Greene \& Young~(\cite{GY92})
\end{table}

\begin{table}
\begin{flushleft}
\caption{ Derived Object Properties}
\vskip 0.1cm
\begin{tabular}{clcccc}
\hline\hline
(1)& (2) & (3)& (4)& (5)& (6)\\
Object&  ST& \Teff  & L$_\star$& \AV & M$_\star$ \\
(ISO\#)&    & (K)&     (L$_\odot$)& (mag)& (M$_J$)   \\
\hline
     &    &      &     &    &      \\
023  &  M7& 2650 & 0.04& 8.0& 30$-$50 \\
030  &  M6& 2700 & 0.07& 3.0& 40$-$80 \\
032  &  M7.5& 2600 & 0.06& 2.0& 30$-$50\\
033  &  M8.5& 2400 & 0.008& 7.0& 8$-$12\\
102  &  M6& 2700 & 0.08& 3.0& 40$-$80\\
160  &  M6& 2700 & 0.04& 6.0& 30$-$60\\
164  &  M6& 2700 & 0.09& 6.0& 40$-$80\\
176  &  M6& 2650 & 0.07& 7.0& 30$-$70\\
\smallskip
193  &  M6& 2650 & 0.1& 7.5& 40$-$80\\
\hline
\hline
\end{tabular}
\end{flushleft}
\end{table}
\label{stars}

\begin{figure*}
\resizebox{\hsize}{!}{\includegraphics{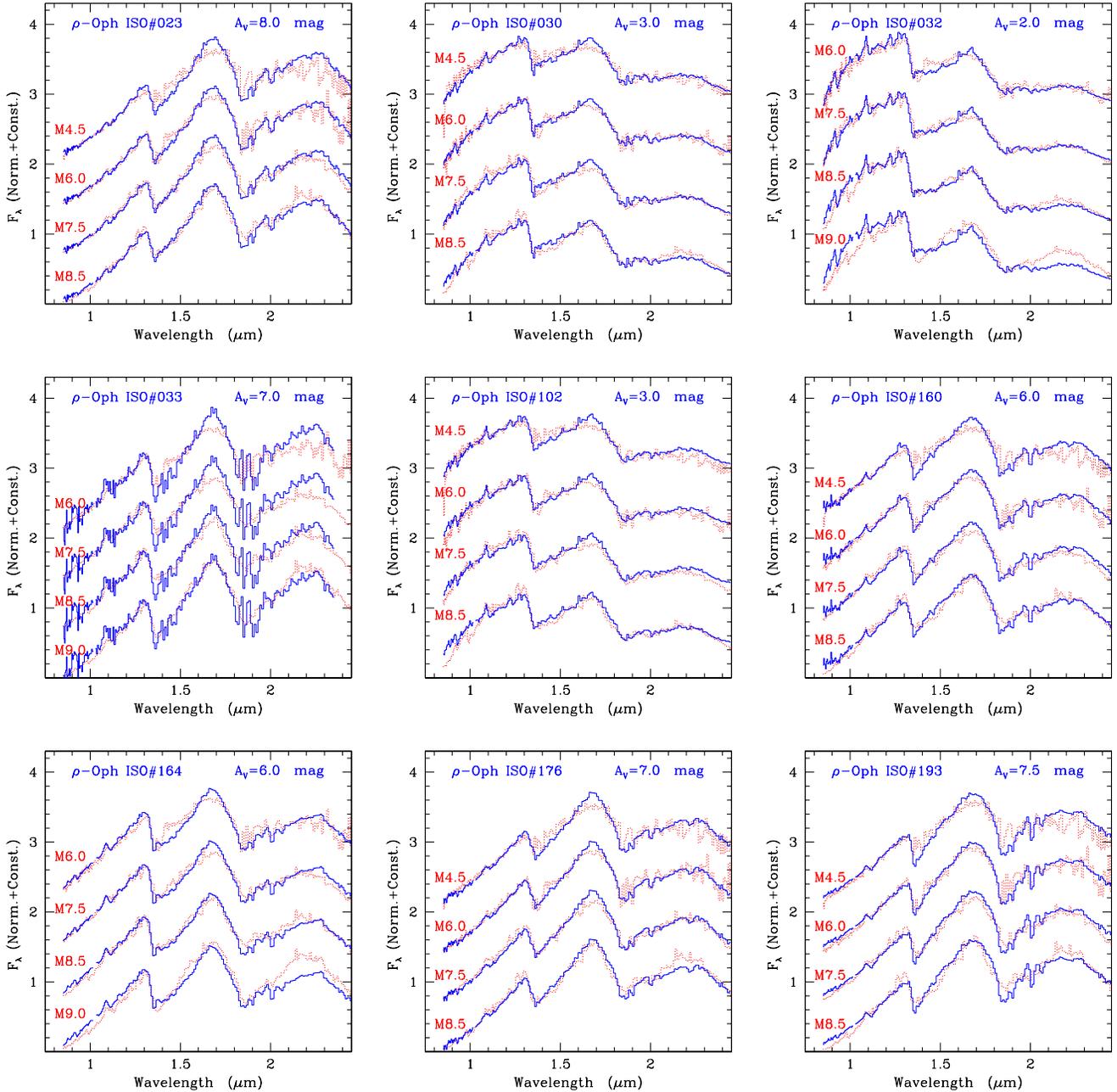}}
\caption{ Observed TNG/Amici spectra of the sample objects.
In each panel, we show (solid  line)
the spectrum of one object compared with the reddened
spectra of field M-dwarfs ( dotted lines) of different spectral types,
from Testi et al.~(\cite{Tea02b}).
All spectra are normalized  to the mean flux in the 1.1--1.75 \um\ range
and shifted with constant offsets for clarity. The field dwarf spectra
have been reddened by the value of \AV\ shown in each panel.
 }
\label{ffield}
\end{figure*}

\begin{figure*}
\resizebox{\hsize}{!}{\includegraphics{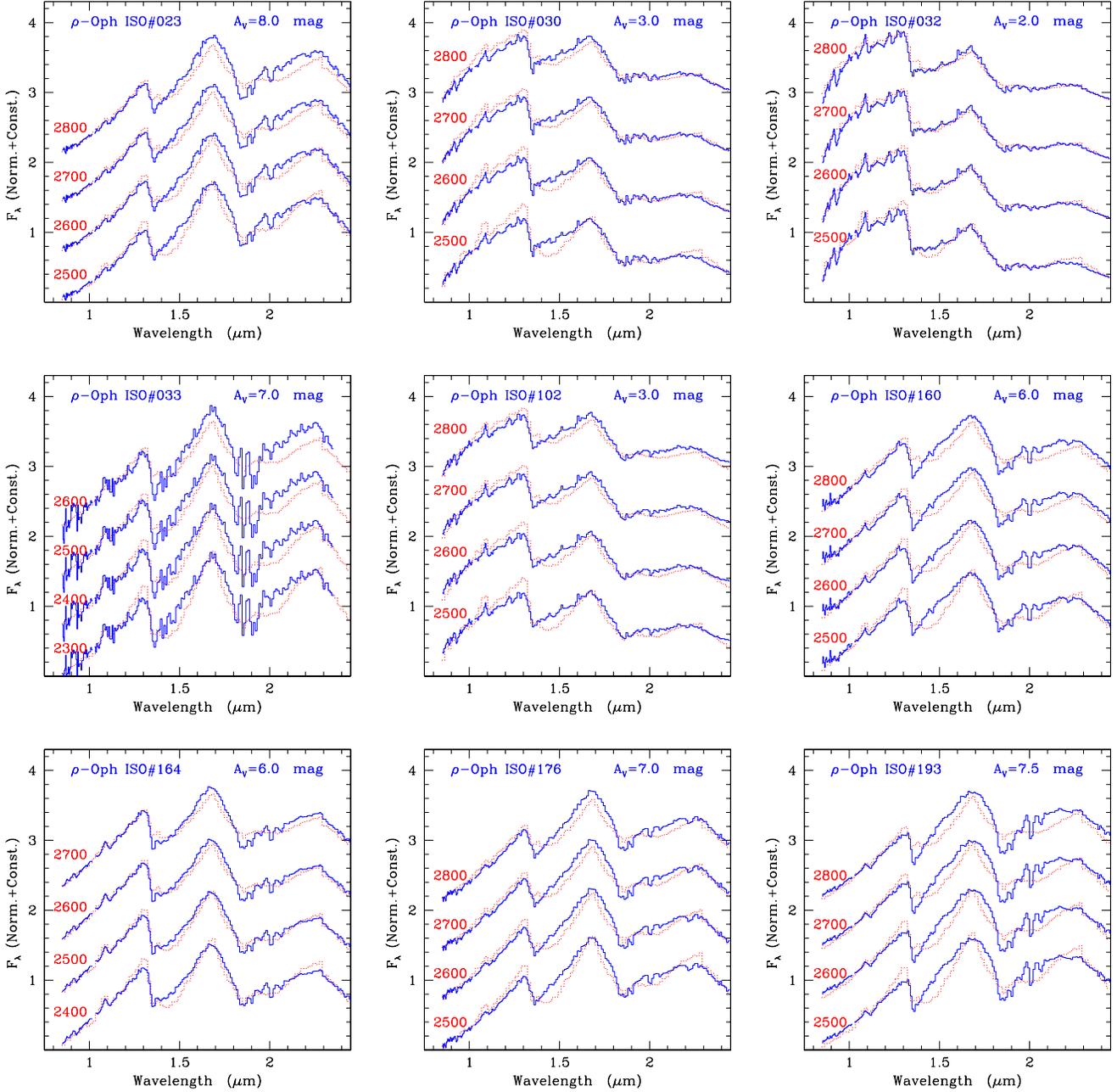}}
\caption{Same as Fig.~\ref{ffield}, but in this case the
red dotted spectra are reddened theoretical atmospheric models
(Allard et al.~\cite{Aea00}), with \Tstar\ as labelled
and log g=3.5.
 }
\label{fatm}
\end{figure*}

The observed
near-infrared spectra obtained at the TNG, normalized to the mean flux
in the interval 1.1 -- 1.75~\um,
are shown in Fig.~\ref{ffield} and ~\ref{fatm}.
We  derive for each object effective temperature and luminosity
in the following manner. We first obtain the extinction and spectral type 
by comparing the source spectra to those of field dwarfs. We then use the
derived extinction value to obtain the effective temperature through the
comparison with reddened model atmospheres. The luminosity is 
computed from the dereddened J-band magnitude using the appropriate
bolometric correction derived from the model atmosphere.

The first step is illustrated in Figure~\ref{ffield}, where we
compare the observed spectra with a set of reddened field dwarfs,
also obtained at the TNG with the same instrumentation
(Testi et al.~\cite{Tea01}, \cite{Tea02b}).
We adopt the extinction law appropriate for \rhooph\
(R=4.2; Cardelli et al.~\cite{CCM89}).

The overall shape of the spectrum from 1 to 2.4\um\ depends strongly
on the spectral type of the object and extinction along the
line of sight. There is, however, 
a degree of degeneracy, so that a cooler, less reddened object looks
similar to a hotter, more reddened one. Therefore, we have also
considered other features, such as the shape of the
H band, the drop due to water absorption at the red edge of the J band and the
intensity of some of the features visible in the J band. 

Although some of  these characteristics depend somewhat on the gravity, so that we 
cannot expect a perfect 
match between  the young BDs and the field dwarfs, the fits are very
good for most objects.
Of the nine targets, three have extinction \AV$\le$ 3 mag,
and six $\ge$6.0 mag; eight  out of nine objects have
a spectral type M6--M7, with $\pm$one subclass uncertainty,
 while \#033 has a later spectral type (M8.5).
We give the results in Table~2, Columns 2 and 5.
Note that even if the extinction is determined in the wavelength range
0.8-2.4 \um, for convenience  we express it in terms
of
\AV , the extinction in the visual,
(\AV = \AJ/0.313; Cardelli et al.~\cite{CCM89}).

In Fig.~\ref{fatm}, the same TNG spectra
 are compared to 
low gravity, $\log g = 3.5$, model
stellar atmospheres (Allard et al.~\cite{Aea00}; \cite{Aea01}), smoothed
to the appropriate resolution and reddened using the
value of A$_V$ derived above.
We obtain from this comparison 
the best value of the effective temperature \Tstar, as well as
a check on the adopted value of \AV.
Our estimates of \Tstar\ have an uncertainty of typically $\pm 100$K;
we assign values of 2600--2700 K to all the objects of spectral type
M6--M7.5, while  \#033 is definitely cooler (\Tstar $\sim$ 2400).
The vales of \Tstar\ are given in Table ~2, Column 3.
The robustness of these results and the uncertainties in the 
derivation of A$_V$ and \Teff\ are described in more detail in Appendix~\ref{a33}.

Since all the objects in our sample  have a mid-infrared excess, we have considered the
possibility that excess emission is present also in the near-infrared,
and is affecting our determination of the stellar parameters. We have
therefore subtracted the maximum contribution expected from an irradiated  disk
 (flared, seen face-on; see \S 4) from  the observed spectra.
We found no significant
change in the derived stellar parameters.

The luminosity of the objects is shown in Table~2,
Column 4. 
It has been computed from the dereddened J flux,
and the ratio of the J to the total flux given by the appropriate
stellar atmosphere model.  These bolometric corrections are virtually identical
to those of Wilking et al.~(\cite {WGM99}) and Leggett et al.~(\cite {Lea02}).
For all stars, the adopted distance is
D=150 pc. The uncertainties on \Lstar\ are difficult
to determine  accurately. We estimate that they probably range
from 20\% to 30\%, mostly due to uncertainties on \AV.
The bolometric correction for the J band changes very little
with the atmospheric parameters,
but an additional uncertainty (the same for all stars) may come from the
uncertainty in the assumed distance.

\begin{figure*}
\resizebox{\hsize}{!}{\includegraphics{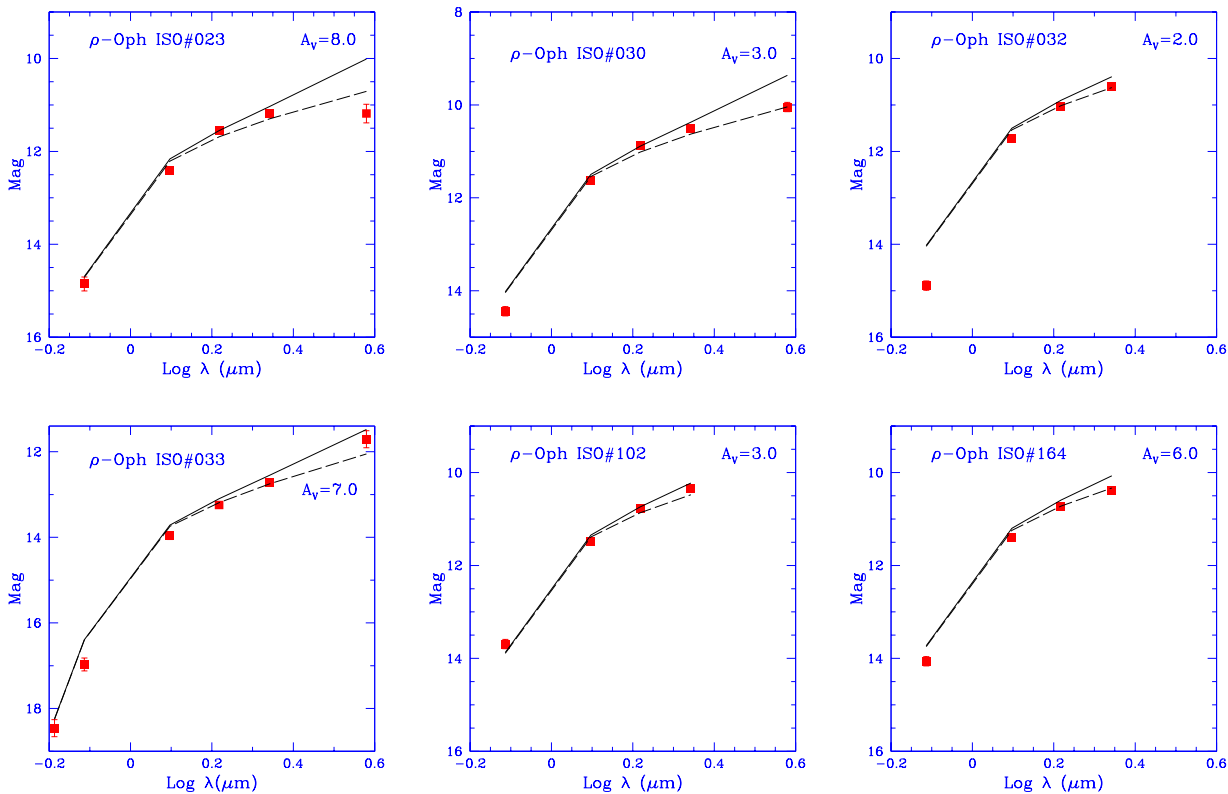}}
\caption{ Comparison between dereddened observed  magnitudes (i, J,H,K, L'),
shown by squares, and the prediction of model atmospheres
with parameters as in Table 2 (dashed lines) and of
the same model atmospheres with additional disk emission (flared, face-on; solid lines)
for the six stars for which i-band magnitudes are available.
Disk models are described in \S 4.
} 
\label{mags}
\end{figure*}

Finally, we have performed a last check on the reliability of our
estimated parameters
using  our i-band photometry.
For each star,
we computed from   model atmospheres
and extinctions  theoretical values of the magnitude in the i, J, H, K
and L
bands by convolving the
flux distribution with the appropriate filter responses. 
The results are shown in Fig.\ref{mags},
where we compare model predictions and observations for the 6 stars
for which i-band photometry was  obtained.
\footnote[1]{ Note that in Paper~II, Fig.4 shows in the inset broad-band fluxes
of \#033 (GY~11) dereddened by \AV =7.5 mag, rather than 7.0 as quoted.}
The agreement of the i-band observed and predicted magnitudes
is generally rather good, given 
the extreme
sensitivity  of the model predictions to the exact shape
of the i-band filter, with the possible exception
of \#032, which would need \AV=3 mag, rather than the 2 mag determined
from the comparison with field dwarfs. The corresponding change in luminosity
would be of 35\%.

\begin{figure*}
\resizebox{\hsize}{!}{\includegraphics{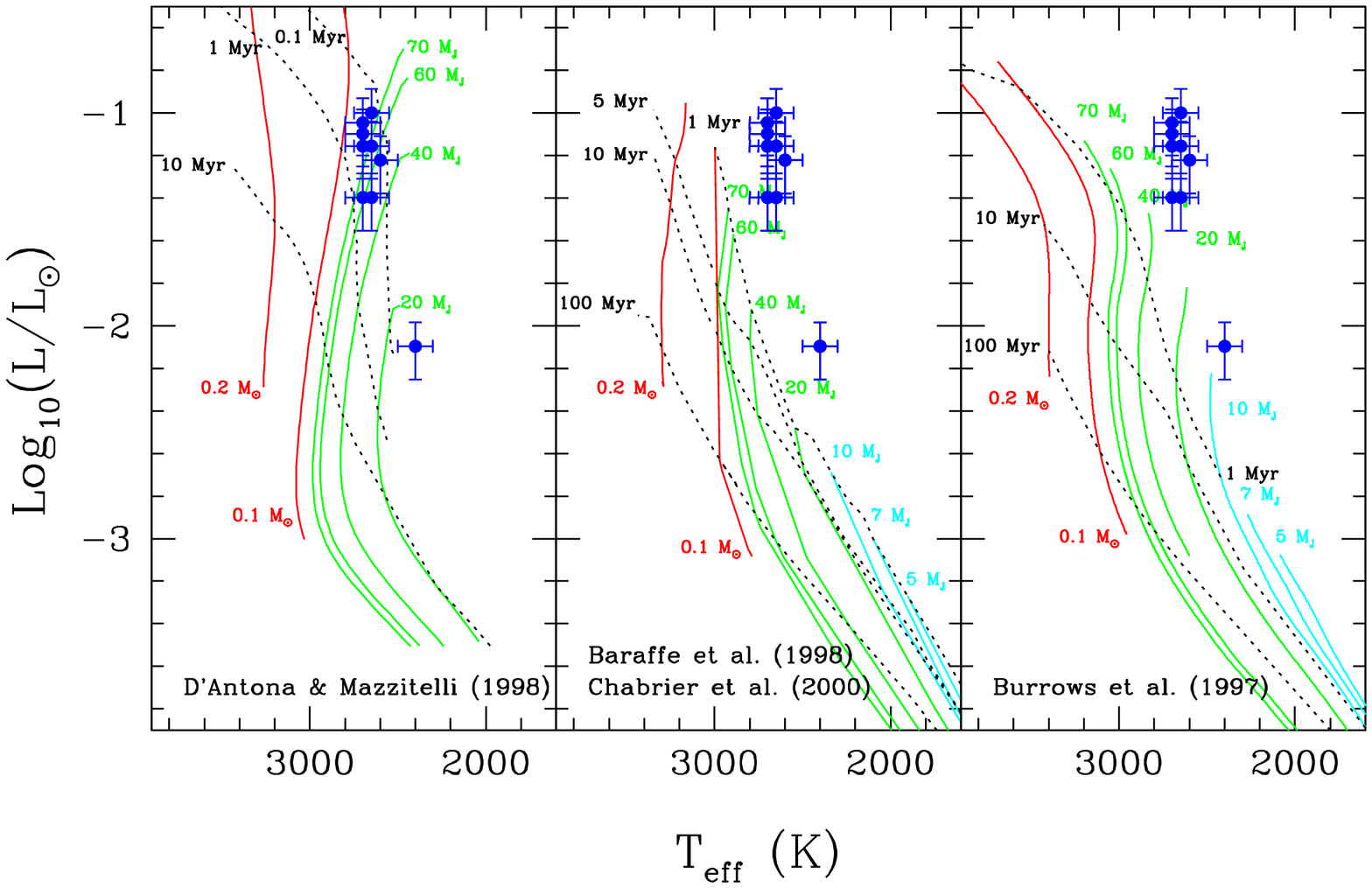}}
\caption{ HR diagram for three sets of evolutionary tracks:
D'Antona \& Mazzitelli~(\cite{DM97}) in the left panel,
Chabrier et al.~(\cite{Chea00}) and Baraffe et al.~(\cite{Bea98}; for the 
0.2 \Msun) in the mid panel, Burrows et al.~(\cite{Bea97}) in the right panel.
Solid lines refer to objects of different mass, as labelled:
hydrogen burning stars in black (red), deuterium burning BDs in 
medium grey (green), objects
below the deuterium burning limit in light grey (cyan). 
Isochrones  are shown as dotted lines, and labelled with the
appropriate age. On each panel, the location
of the nine observed objects is shown by dots with error bars.
 }
\label{HR}
\end{figure*}

Fig.~\ref{HR} shows the location of the nine \rhooph\ objects
in the HR diagram. In the three panels, we overlay them to three
different sets of evolutionary tracks, computed by D'Antona
\& Mazzitelli~(\cite{DM97}), Chabrier et al.~(\cite{Chea00})
 and Burrows et al.~(\cite{Bea97}),
respectively.  The derived masses (Table~2, Column 6)
depend on the adopted tracks, hence we report the corresponding
range of values.
All objects appear to be very young, with ages 
lower than 1 Myr and probably of the order of a few $10^5$ yr.
It is well known that at such ages evolutionary tracks are not very reliable
(Baraffe et al.~\cite{Bar02}),
and that the parameters derived from the location on the HR diagram are only
indicative. 
However,
in spite of the uncertainties in both tracks and
observations,   we   estimate that
our sample contains one  very low mass object
(\#033), with a mass of only $\sim$8--12~\MJ\ (Paper~II), and a  group of
objects with masses in the BD range, of which about half (\#023, \#032, \#160
and \#176) are very likely  BDs.

The clustering of eight out of nine of our objects in a narrow region of
the HR diagram is a
result of our selection criteria  and can be understood as follows.
TTS in \rhooph\ have typical
ages of 1 Myr, with very few stars as old as 3 Myr (Palla \& Stahler~\cite{PS00}). 
The lack of older BDs in the sample is easy to understand, since
the limited sensitivity of the 
ISOCAM survey (especially at 14.3 \um)
strongly biases towards the highest luminosity, hence the youngest
sources, and we  expect to find in our sample 
only   BDs younger than the average
TTS. Older, more massive objects  could, in principle,  fall
in our sample.
In practice, we found that this was not the case, given \rhooph\ 
typical age.
We have applied the procedure adopted by
Bontemps et al.~(\cite{Bonea01}) to a ``theoretical"
star with mass 0.2 \Msun\ and age of 2 Myr,
using model-predicted J, H, K magnitudes (Baraffe et al.~\cite{Bea98})
and \AV $\leq$9 mag; 
such a star would have 
a computed luminosity higher than our selected
upper limit (\Lstar$\simless$ 0.04 \Lsun), and would therefore not
be included in our sample.  Younger stars would be even more luminous.

\section {Disk models}

\begin{figure*}
\resizebox{\hsize}{!}{\includegraphics{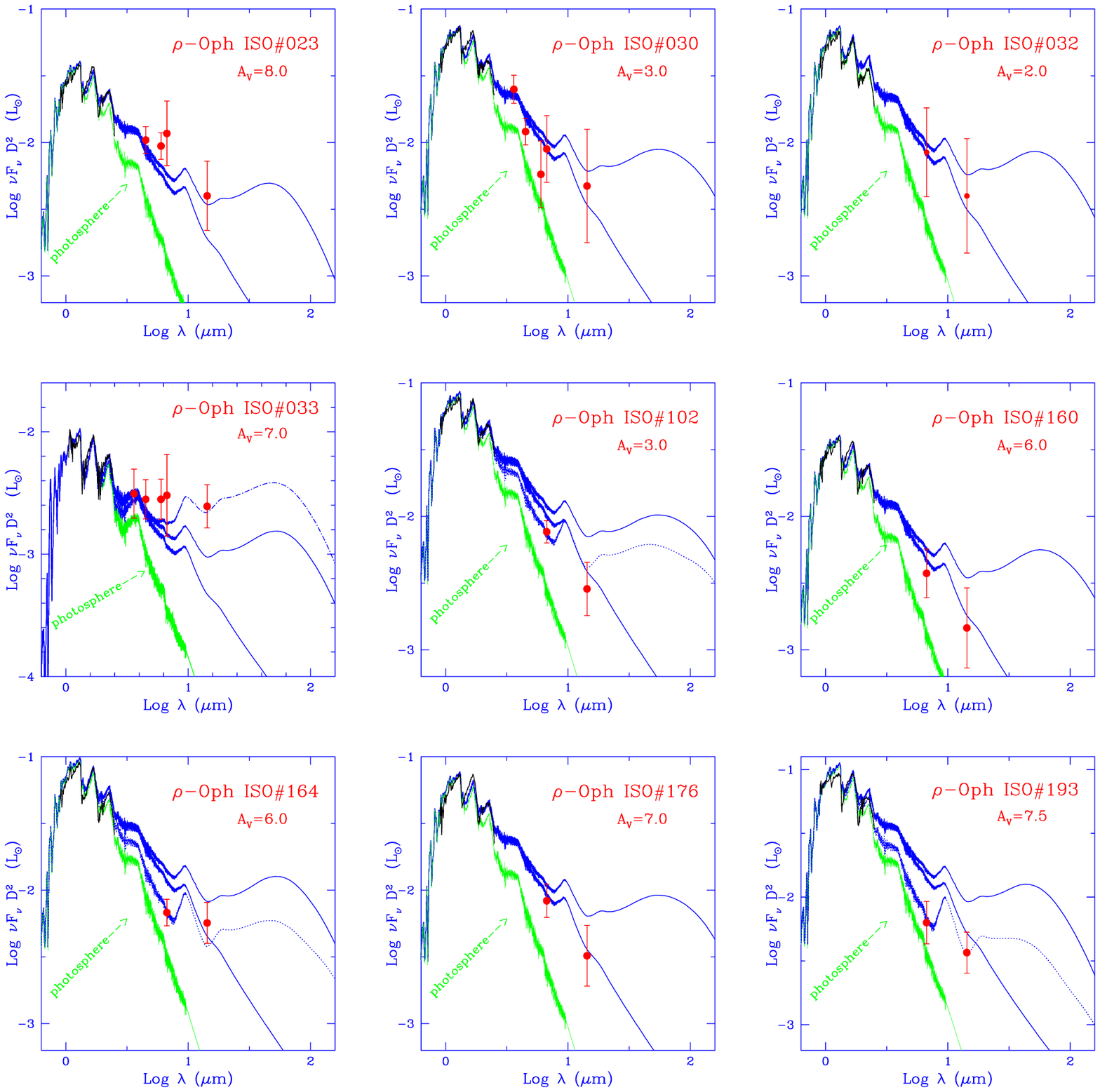}}

\caption{ Disk and photosphere predicted SEDs. In each panel,
the red dots with error bars show the ISOCAM observed fluxes
(Comer\'on et al.~\cite{Comea98}; Bontemps et al.~\cite{Bonea01}).
The black solid line is the dereddened and calibrated TNG/Amici spectrum.
The green jagged line shows the SED of the  photosphere. The combined SED
of the photosphere plus disk is shown by blue lines; in each panel,
the two solid curves refer to face-on flared  (upper curve) and flat disks
(lower curve), with \Rin=\Rstar.
For \#033, the dot-dashed curve shows the SED of a face-on, flared disk
with \Rin=3\Rstar.
Finally, we show on three panels the SEDs 
of tilted flared disks (dotted lines), seen by the observer with inclinations of
69$^o$ (\#102), 80$^o$ (\#164) and 86$^o$ (\#193) respectively
(0$^o$ for face-on disks).
 }
\label{seds}
\end{figure*}

All nine  objects  have  mid-infrared fluxes measured with ISOCAM
in at least two
bands (centered at 6.7 and 14.3 \um; Bontemps et al.~\cite{Bonea01}). In three cases,
there are additional ISOCAM observations in three narrower bands,
centered at 3.6, 4.5 and 6.0 \um; Comer\'on et al.~\cite{Comea98}).
The ISOCAM points are shown for each object in Fig.~\ref{seds}, together with
our calibrated and de-reddened TNG spectra.

For each system we  compute the SED predicted by disk models, 
assuming that the disk is heated by the radiation of the central object.
We ignore in this paper any possible viscous heating within the disk (see \S 6 for a 
brief discussion).
We follow as in Paper~I and~II the method
outlined by Chiang \& Goldreich~(\cite{CG97}; CG97), with some improvements and
modifications  (Natta et al.~\cite{Nea01}; Chiang et al.~\cite{Cea01}). 
CG97 consider a disk in hydrostatic equilibrium in the vertical direction
(flared), and describe at each radius 
the vertical temperature structure of the disk in
terms of two components:
the disk surface, i.e., the external layer of the
disk which is optically thin to the stellar radiation, and the disk midplane.
These models allow a quick and reasonably accurate description of the expected SED,
more than adequate for the purposes of this paper.

The disk is a scaled-down version of TTS typical disks. It extends inward to the stellar
radius, and outward to \Rd =$1\times 10^{15}$ cm (67 AU).
The total mass is \Md $\sim$0.03 \Mstar, and the surface
density varies as R$^{-1}$. The dust in the disk midplane has opacity
$\kappa=0.01 (\lambda/1.3{\rm mm})^{-1}$ cm$^2$ g$^{-1}$ (Beckwith et al.~\cite{BSCG90}).
For the dust on the disk surface, we take the
mixture of carbonaceous materials and silicates that provides a good fit to the SEDs
of several \pms\ stars (Natta et al.~\cite{Nea01}), i.e., a MRN distribution of graphite
and astronomical silicates with $dn/da \propto a^{-3.5}$, $a_{min}$=100 \AA,
$a_{max}$=1 \um, 30\% of cosmic C and all Si into grains.

The results of the model calculations are shown in Fig.~\ref{seds}.
The stellar parameters (\Tstar, \Lstar, \Mstar) are taken from
Table~2. 
As pointed out in Paper I, most of the disk
parameters are irrelevant for the calculation of the mid-infrared disk emission,
or appear in combinations, and cannot be determined individually 
(see also Chiang et al.~\cite{Cea01}). 
As long as the disk midplane remains optically thick
to mid-infrared radiation,
the only parameters that affect the SED  in the near and mid-infrared
are the geometrical shape of the disk (i.e., the flaring angle), the
inclination to the line of sight and, to some degree,
the disk inner radius $R_i$. 
There is also some dependence of the shape of the SED on
the surface dust model; however, since the luminosity intercepted and
re-radiated by the optically thin surface layers
is fixed, variations due to (reasonable) changes of the
grain properties  are well within the uncertainty of the existing observations.

 The upper solid curves in Fig.~\ref{seds} show 
the SEDs of flared disks with $R_i$=\Rstar, seen face-on.
They all have
strong silicate emission at 10 \um\ and a rather flat spectral slope between the two
ISO bands at 6.7 and 14.3 \um,  of order $\alpha \sim 0.6-0.8$ ($\nu F_\nu \propto \nu^\alpha$).
If, rather than extending all the way to the
stellar surface the disk is truncated further out, as predicted by magnetospheric 
accretion models in TTS, at each radius
the surface of a flared disk intercepts and reprocesses a larger
fraction of the stellar radiation.
The disk emission increases correspondingly  at
all wavelengths but in the near-infrared, where one is sensitive to
the lack of the hottest disk dust.  A  model with $R_i\sim$ 3\Rstar\
is shown (dashed line) for \#033, where, as discussed in Paper II, the
inner hole may account for the large
observed mid-infrared excess. 

Large variations of the predicted SED occur if the disk shape changes.
On each panel, we  show the predictions of geometrically thin,
``flat" disks (lower solid lines), i.e., disks where the grains are not
well mixed with the gas, but have collapsed  onto the disk midplane.
Also for these models, we have adopted the CG97 formalism, which remains
adequate in all the cases where the disk heating is dominated by
the stellar irradiation. If the surface contribution to the SED were
negligible, one would recover for these disks the well known temperature
profile $T\propto R^{-3/4}$ and the power-law slope of the SED $\nu F_\nu \propto \nu^{4/3}$ (Adams and Shu \cite{AS86}).
Our calculations 
show that also in flat disks the surface contributes 
to the
mid-infrared flux, as shown by the presence in the SED of the silicate feature
in emission; however, the midplane emission is larger than the surface contribution
at all wavelengths but  in the region $\sim$ 8--12 \um,
where the silicate feature dominates,
so that  the spectral slope between the two ISO points  is  always very
close to 4/3.
At all wavelengths larger than $\sim$ 2.5 \um,
the emission of a flat disk is
significantly lower than that of a flared one.

Finally, we show on three Panels of Fig.~\ref{seds} the predictions
of tilted flared disks, seen by the observer with inclinations of
69$^o$ (\#102), 80$^o$ (\#164) and 86$^o$ (\#193) respectively
(0$^o$ for face-on disks).

The comparison of the ISO observations to the model predictions 
shows  that irradiated disk models can account  for the observed
mid-infrared excess. More precisely, and in spite of the large uncertainties
 of the ISO data,
inspection of Fig.~\ref{seds} shows that there are five stars out of nine
(\#030, \#032, \#102, \#160, \#176) that are extremely well fit by flat disk models.
Two objects (\#023 and \#033) seem to require flared, face-on disks, while two
others have a lower mid-infrared excess, consistent with disks 
seen  rather edge-on.
However, given the large error bars and the model uncertainties, most
objects with flat disks are also consistent with flared disk models
with large inclination, as shown for the case of \#102.


\section {Discussion}

\subsection {Photospheric Parameters}
\label{sdiscspar}

The agreement between the object spectra and those of field dwarfs is good beyond our
expectations.  The largest differences are of 20\% at most, generally at the peak 
of the H band, with no systematic difference between objects with large or low extinction,
nor between stronger and weaker sources.
Even the relatively narrow features that appear in the spectra around 
1.1 \um\ are
often well matched in the two sets of spectra.
This indicates that, at the resolution of our observations, one should not expect
strong gravity effects. We have checked that this is indeed the case by
comparing model atmosphere spectra smoothed to the observed resolution
for stars of different gravity (Allard et al.~\cite{Aea01}), ranging from 3.5 to 6.0.
All the models with gravity in the interval 3.5--5.0 are 
practically identical, at our spectral resolution and in this temperature range.

The comparison of our spectra with model atmosphere predictions is somewhat less
satisfactory, especially in the  H band, where the shape of the
feature  peaked at about 1.7 \um\ (resulting from water absorption
features at shorter and longer wavelengths) is narrower in the models than
observed, and around 1.3 \um, where the models tend to predict
more emission than is observed. Note, however, that this is not always the case (see,
for example, \#032 and \#164). Still, the agreement is in general rather good,
with differences that never reach more 30\%, again with no dependence on the
extinction nor on the observed signal.

The comparison of our determinations of the photospheric parameters
of individual
objects with previous spectroscopic determinations in the literature
shows that in some cases there is good agreement, while in others there
are discrepancies that are not easily understood. For example,
Wilking et al.~(\cite{WGM99}) assign similar spectral types to
\#023, \#030, but a significantly later one (M8.5)
to \#164, based on K band 
R$\sim$300 spectroscopy. For the same object, Luhman \& Rieke ~(\cite{LR99})
estimate a spectral type M7, based on intermediate resolution K band spectroscopy, similar to our classification M6.
The same authors, on the other hand,
attribute to \#030 a somewhat earlier spectral type (M5-M6).
The case of \#033 (GY~11) has been discussed in detail in Paper~II.
A likely reason for differences in the spectral classification
is that our scheme is based on the overall spectral shape, while the others
rely on fitting individual spectral features, which in the infrared show
large scatter for late M objects (e.g., Luhman \& Rieke~\cite{LR98}).

On the more general issue of the effective temperature
scale of  young BDs, we attribute 
temperatures in the range 2600-2700 K to our group of objects
with spectral types M6--M7.5. Our only object with later
spectral type (M8.5) has \Tstar =2400$\pm$100 K. 
In a preliminary analysis of our sample 
field dwarfs (Testi et al.~\cite{Tea02b}), we derive a similar
effective temperature-spectral type correspondence.
This is not significantly different from the scale used by
Wilking et al.~(\cite{WGM99}) in their study of candidate BDs in \rhooph. 
It is, however, at odds with
some recent results, that tend to attribute to young BDs of similar spectral
types temperatures
higher than our values (Lucas et al.~ \cite{Lucea01};
Lodieu et al. ~\cite{Lodea02}).  
Further work, on larger samples of BDs in young star forming regions
is clearly required.

\subsection{ The disk hypothesis}
 
The comparison between  models and observations, discussed in the
previous section, proves that the mid-infrared excess
associated to many young BDs can be accounted for by the emission
of circumstellar disks heated by the radiation of the central object.

Few disk properties are  constrained by the existing observations,
and we do not want to overinterpret our results,
given the large uncertainties of the observed fluxes,
and the simplicity of the adopted models.
However, 
in our limited sample of nine stars 
we find disks of different flavours, and, in particular,
an indication that many BDs  may
have flat disks. 
If we consider also the three objects in Cha I studied in Paper I,
we have three objects with clear evidence of flared disks, and nine where flat
disks seem more appropriate, although we cannot  rule out
almost edge-on flared disks for some of them (see also Apai et al.~\cite{apai02}). 
This is  potentially an interesting
result, since it seems natural to associate flat disks with  
dust sedimentation toward the midplane.
In our selection of ISO sources, we have an obvious strong bias against
objects with flat disks, since we required that the sources  were
detected by ISO in both bands. So, the fact that our objects with the
lowest 6.7 \um\ fluxes (\1\ and \#033) have flared disks is not surprising.
However, there is   no bias against selecting flared disk objects
of higher luminosity, and we find   only one (\#023).
The possibility of dust settling in these very young low-mass objects
is intriguing. However, it needs to be confirmed by high-quality
photometric
observations at longer wavelengths, before entering into further
speculations. 

The ejected embryos hypothesis does not exclude that BDs may have a small,
and therefore short-lived, circumstellar disk. Estimates by Bate et
al.~(\cite{Bate02})
give  disk radii of about 20 AU or less. The existing infrared
data do not allow us to rule out such possibility, since the SED of a model
with \Rd=20 AU will differ from the SED of a disk with \Rd=75 AU only at wavelengths
$\simgreat$ 40 \um. The mass of the disk is not predicted by
the Bate et al.~(\cite{Bate02}) calculations, nor constrained by the existing
observations,
since the only constraint we can set 
is that the disk has to be optically thick in the mid-infrared. This,
however, only requires a disk mass of 10$^{-5}$--10$^{-6}$ \Msun
(or \Md/Mstar$\sim 10^{-4}$), which is still
consistent with a typical disk (having \Md/Mstar$\sim$ 0.03, \Rd=75 AU), truncated at
\Rd=20 AU. Until far-infrared and millimeter data  become available,
the only way to validate these models is to determine the fraction
of disks in  unbiased samples of BDs of known age.

Finally, one should remember that our analysis
relies on the assumption that the ISO
sources coincide with the objects we identify in the near-infrared.
In some cases, this is likely to be true (see Appendix~\ref{images}
and the discussion of \#033 in Paper~II).
In other cases,
it is impossible  to check the validity of this assumption,
given the large ISO beam and the presence of other red objects
in the near-infrared images. However, the good agreement between the
observations and the model predictions, which depend essentially only
on the stellar properties we derive from the spectroscopy, is encouraging. 
Further tests of the association of the observed mid-infrared excess with the
identified stars could be obtained by accurate images 
in the L and M bands, where we predict that the disk emission should be
dominant (see Fig.~\ref{seds}).



\section {Conclusions}

We have discussed in this paper  a sample of nine very low-mass
objects in the \rhooph\ star forming region that have 
evidence for circumstellar warm dust. We selected from the ISOCAM 
sample of Bontemps et al.~(\cite{Bonea01})
those objects
that have mid-infrared detections in both the 6.7 and the 14.3 \um\ bands,
relatively low extinction and low luminosity.
We determined first
if these BD candidates
were indeed bona-fide BDs, and then we checked if the observed infrared excess was
consistent with the predictions of disk models, similar in properties to those
associated to T Tauri stars.

Our strategy was very successfull. The low-resolution near-infrared
spectra obtained at the TNG
 allowed us to determine  for each object spectral type and
extinction, by comparison with field dwarfs observed with the same
instrumental set-up, as well as  effective temperature and luminosity, by
comparison with model atmosphere predictions. The comparison with various
sets of evolutionary tracks on the HR diagram shows that all the nine sources are
very young, low-mass objects. 
In particular, one (\#033 or GY 11, already discussed in Paper II) has a mass
of 8-12 \MJ, while the others have masses in the BD mass range; four of them are very
likely bona-fide BDs.

In all objects, the mid-infrared excess  is consistent with the predictions of
disks irradiated by the central object. 
We find no evidence of strong accretion occurring in these systems,
based on the fact the
observed near-infrared fluxes are dominated by the emission of the
photospheres, and there is very little contribution (if any) from
hot dust. However, it is not clear to which degree 
the near-infrared excess  in  very low-luminosity objects
is a sensitive indicator of accretion
(see, for an example of an actively accreting object with no near-infrared
excess, Fern\'andez \& Comer\'on \cite{FC01}), and this issue should be explored more
quantitatively in the future.

The existing data indicate that the disks must be optically thick
at mid-infrared wavelengths; in some cases they must be flared
(i.e., gas-rich with well-mixed dust and gas), while in others it is
possible that they are geometrically flat, i.e., that dust has settled to
the disk midplane. However, data at longer wavelengths are necessary to further
investigate this point, and we do not want to put too much weight on this
rather weak evidence.

In the same sobering vein, we want to point out that our results do not  discriminate
yet between  different formation mechanisms, namely between the possibility
that BDs form from the gravitational collapse of individual, very low-mass cores, and
the ejected embryo theory. 
We fit the observed mid-infrared excess with 
a scaled-down version of disks around the more massive TTS.  This,
however,  just implies   that ``normal" disks can
account for the existing observations, since few parameters are
actually constrained. As already pointed out
in Paper I and II, only observations at long
wavelengths can measure the disk radius and
mass, since the lower limits that we can derive
from the conditions that the disk is optically thick in the mid-infrared
are hardly significant.

Having stressed all the limitations of our results, let us now point out that
this is the first sample of very low mass objects in a star forming regions
where evidence for circumstellar disks has been found and
investigated in detail.
Our accurate near-infrared spectroscopy, which allows us to estimate a reliable
value of the mass of the objects, proves  that
disks exist around  low mass objects, well into the range of
brown dwarfs. In one case, \#033, our data provide strong indications that
an object with mass close to or below the deuterium burning limit also
has a circumstellar disk.  In addition to providing the beginning of
a census of disk properties around BDs, our models  indicate that the
excess due to the cold disks irradiated by a central BD can only be detected
by deep photometry in the L and M bands. 
We expect that  major progress in our
understanding of BD formation will be obtained
by combining near-infrared low resolution spectroscopy
with photometry in J,H,K,L,M of unbiased (i.e., not a-priori selected because
they have a mid-infrared excess, as here) samples of BD candidates in star forming
regions of different age.

\begin{acknowledgements}
We thank Carsten Dominik and Michael Meyer for useful discussions.
It is a pleasure to acknowledge the TNG and ESO staff for their
excellent support during observations.
This publication makes use of data products from the Two Micron All Sky Survey, which is a joint project of the University of Massachusetts and the Infrared Processing and Analysis Center/California Institute of Technology,
funded by the National Aeronautics and Space Administration and the National Science Foundation.
This work was partly supported by  ASI grant  ARS 1/R/27/00 to the
Osservatorio di Arcetri.
\end{acknowledgements}

\appendix

\section{K$_s$ or K' finding charts}
\label{images}
\begin{figure*}
\resizebox{\hsize}{!}{\includegraphics{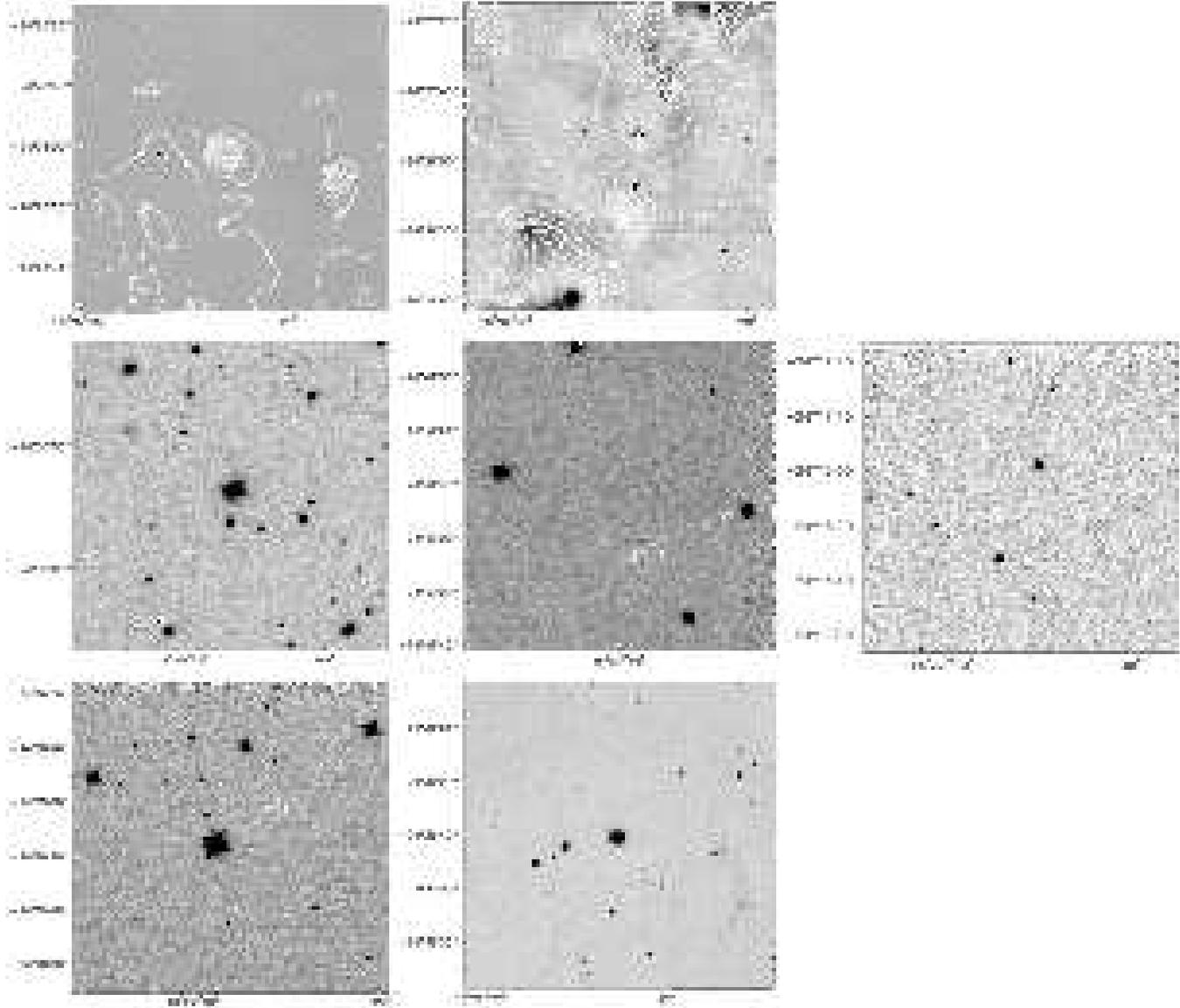}}
\caption{ Finding charts for eight of the nine sources: 2.2 \um\ images are shown
as grey scale, ISOCAM-LW1 images are shown as contour plots. See text for
details.}
\label{fcharts}
\end{figure*}

Fig.~\ref{fcharts} shows 2.2 \um\ finding charts for eight of the nine sources
discussed in this paper. The grey scale images have been obtained at the
TNG with NICS (\#023, \#030, \#032) by us on July 2001,
at ESO/UT1 with ISAAC (\#033) and at ESO/NTT with SOFI (\#102, \#160, \#164, \#176).
The ISAAC and SOFI data have been extracted from the ESO Science Archive;
they have been originally obtained for ESO proposals 63.I--0691, 
65.I--0576,  67.C--0325 and 67.C--0349.
Finding chart for \#193 can be obtained from the  2MASS database at
http://irsa.ipac.caltech.edu.
For sources \#023, \#030 and \#033, contours show
the ISOCAM-LW1 emission (Comer\'on et al.~\cite{Comea98}).

\section{Accuracy in the determination of the photospheric parameters:
A$_V$, Spectral Type, and T$_{\rm{eff}}$}
\label{a33}

In this appendix we discuss the accuracy of the method used to derive 
extinction, spectral types and effective temperatures from the low-resolution
near infrared spectra. For this purpose we will discuss the derivation of the 
parameters for three extreme cases, the source with the highest 
extinction (\#023), the source with the lowest extinction (\#032), and the
later spectral type source (\#033).

The procedure starts with the determination of the extinction and spectral 
type based on the comparison with field dwarfs spectra obtained with the same
instrumental setup (Testi et al.~\cite{Tea01}; \cite{Tea02b}). For each source
we compare the observed spectrum with those of the field dwarfs reddened 
by different amounts, the value of the extinction is varied until the 
best match is found with some of the field dwarfs. In practice, we varied the
value of A$_V$ in steps of 0.5 mags, since smaller variations cannot be
significantly distinguished in the comparison. 

\begin{figure*}
\resizebox{\hsize}{!}{\includegraphics{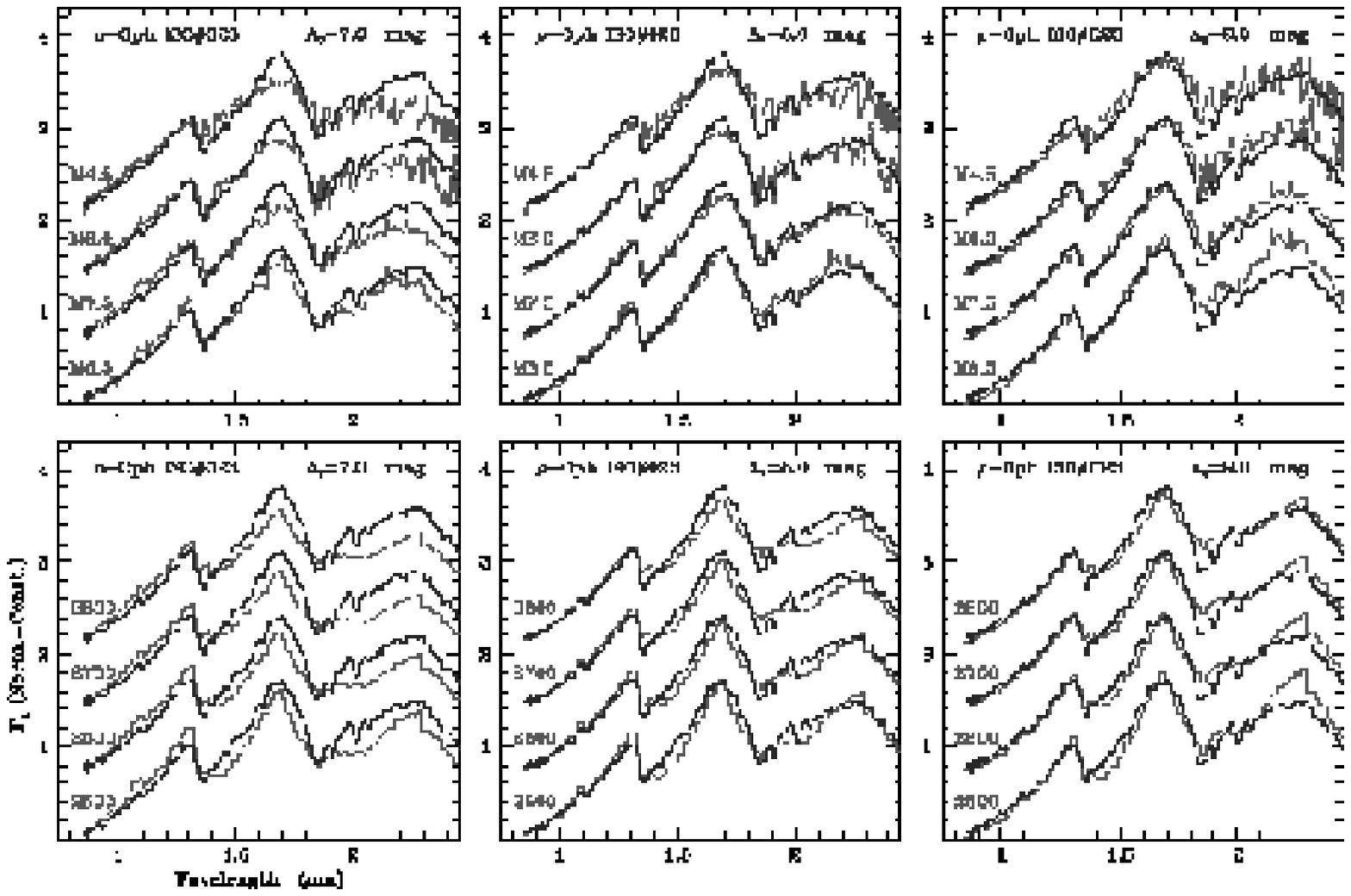}}
\caption{Top panels: the observed spectrum of source \#023 (thick line) is
compared with reddened spectra of field dwarfs (dotted lines); the values of
the extinction used are A$_V=$7, 8, and 9, as labelled on each panel. Bottom
panels: the observed spectrum of \#023 (thick line) is compared with reddened
model atmospheres (dotted lines). All spectra are normalized in the region 
1.1--1.75~$\mu$m and scaled for clarity.}
\label{fa23}
\end{figure*}
In the top panels of Figure~\ref{fa23} we show this comparison for source \#023
and for three values of A$_V=$7, 8, and 9. The best matches, based on the
shapes of the J and H bands, and the fact that the source spectrum at K
cannot be lower than the field dwarfs, to allow for a possible contribution from
disk emission,
are found for A$_V=$8~mags and spectral types between M6 and M7.5. 

Once the spectral type and extinction have been estimated, the source spectrum
is compared with appropriate surface gravity (Log(g)=3.5) model atmospheres
(Allard et al.~\cite{Aea00}), which provide
an estimate for the effective temperature and an additional check on the
extinction. In the bottom panels of Figure~\ref{fa23} we present this
comparison for source \#023. The comparison with the atmospheres confirms that
the estimate of A$_V=$8~mags provides the best match, and show that the two 
models that match more closely the observed spectrum are those with temperature
2600 and 2700~K, hence the best estimate that we derive is \Teff =2650~K,
with an uncertainty of $\sim 100$~K.

\begin{figure*}
\resizebox{\hsize}{!}{\includegraphics{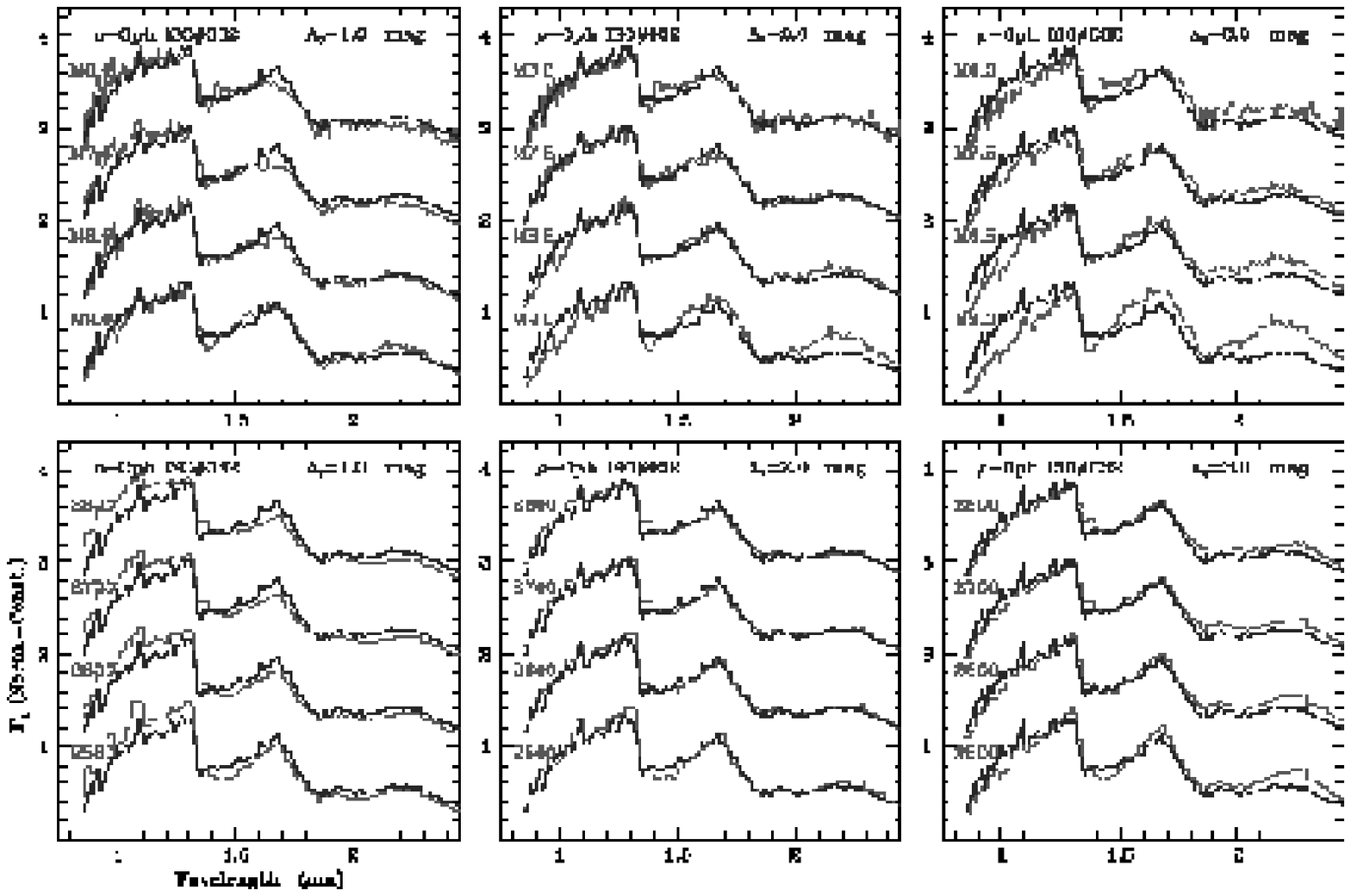}}
\caption{Same as Figure~\ref{fa23}, but for source \#032, the values of the 
extinction in this case are: A$_V=$1, 2, and 3~mags.}
\label{fa32}
\end{figure*}
In Figure~\ref{fa32} we show the procedure for the source in our sample 
with the lowest value of the extinction (\#032). Given the lower extinction, the
spectrum clearly shows some features in the J-band that correspond to the
combination of several blended photospheric absorption lines and bands.
Also in this case we show the comparison with field dwarfs for three values of
A$_V$ spaced by 1 magnitude. In this case, it is possible to obtain a good 
match with a later spectral type and a lower extinction (A$_V=1$, M8.5); however,
the match with M7.5 and A$_V=$2~mags is the best. 
This choice of the extinction is confirmed by the comparison with the model 
atmospheres in the bottom panels of Fig.~\ref{fa32}. The effective temperature 
is estimated to be 2600$\pm$100~K, as for the previous source.

\begin{figure*}
\resizebox{\hsize}{!}{\includegraphics{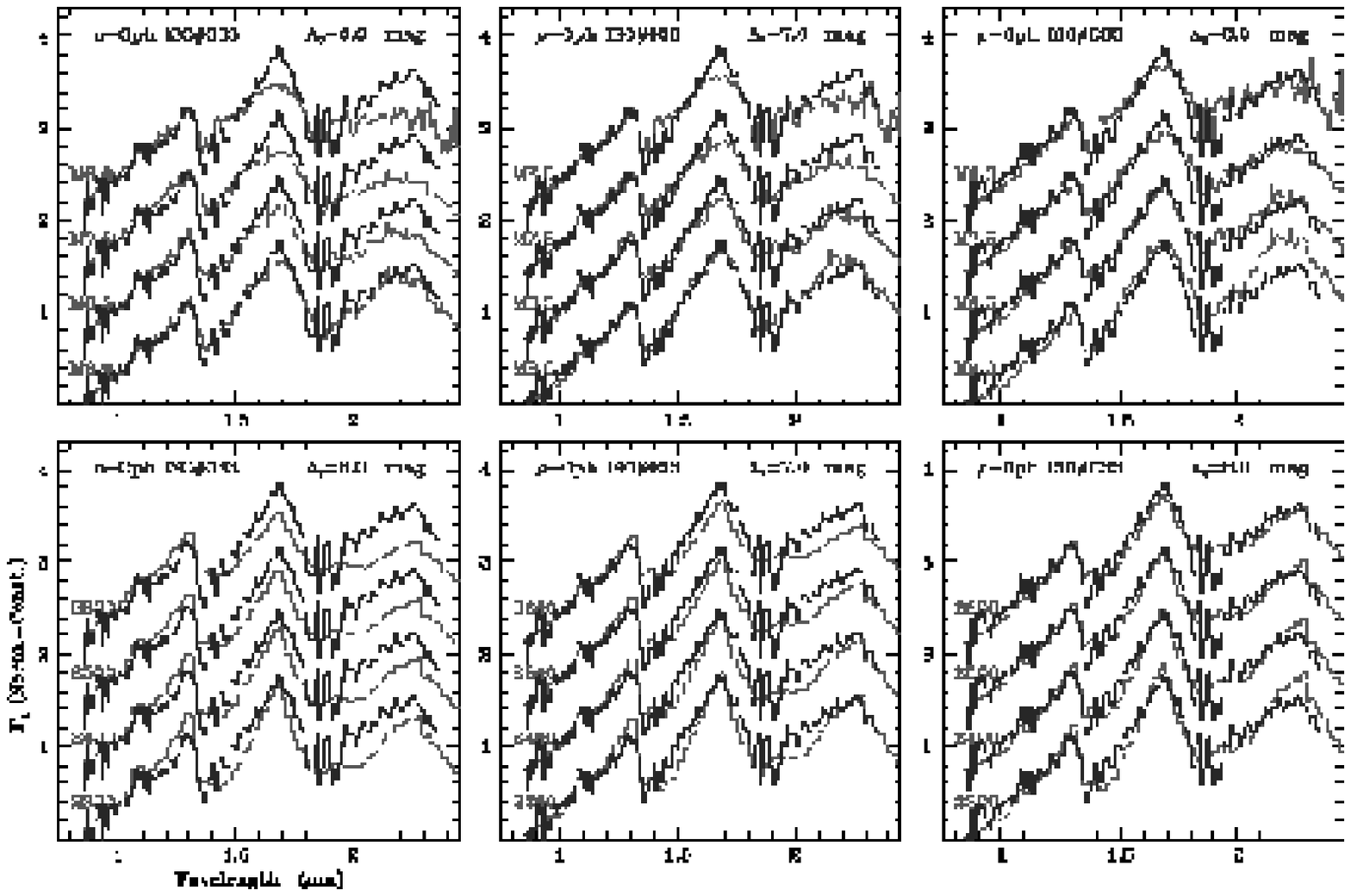}}
\caption{Same as Figure~\ref{fa23}, but for source \#033, the values of the 
extinction in this case are: A$_V=$6, 7, and 8~mags.}
\label{fa33}
\end{figure*}
The source for which the derivation of the parameters is most complicated is
\#033 (Fig.~\ref{fa33}) because of the high extinction and the lower signal
to noise of the spectrum. In this case good matches can be obtained for the
lower extinction and later spectral type, A$_V=$6~mag M9.0, as well as higher
extinction and earlier spectral type, A$_V=$8~mag M7.5. The comparison with
model atmospheres would favor the higher extinction. Thus, also
considering the
uncertainties in the model atmospheres at these spectral types
(see e.g. Leggett et al.~\cite{Lea01}), our best estimates 
are: A$_V=$7.5$\pm$1~mag, M8.5 with one subclass uncertainty and
\Teff =2400$\pm$100~K.


\newpage
\begin{thebibliography}{}

\bibitem[1986]{AS86}
Adams F.C.,  Shu F.H. 1986, ApJ, 308, 836

\bibitem[2000]{Aea00}
  Allard F., Haushildt P.H.,  Schweitzer A. 2000, ApJ, 539, 366

\bibitem[2001]{Aea01}
  Allard F., Haushildt P.H., Alexander D.R., Tamanai A.,  Schweitzer A.
   2001, ApJ, 556, 357
\bibitem[2002]{apai02}
Apai I., Pascucci I., Henning Th., et al. 2002, ApJ, submitted

\bibitem[2001]{Bea01}
  Baffa C., et al. 2001, A\&A 378, 722
\bibitem[2002]{Bar02}
  Baraffe I., Chabrier G., Allard F.,   Hauschildt P.H. 2002, A\&A, 382, 563

\bibitem[1998]{Bea98}
  Baraffe I., Chabrier G., Allard F.,   Hauschildt P.H. 1998, A\&A, 337, 403


\bibitem[2002]{Bate02}
Bate M.R., Bonnell I.A.,    Bromm V. 2002, MNRAS, 332, L65

\bibitem[1990]{BSCG90}
  Beckwith S.V.W., Sargent A.I., Chini R.S.,   Guesten R. 1990, AJ, 99, 924

\bibitem[2001]{Bonea01}
Bontemps S., Andr\'e P., Kaas A.A., et al. 2001, A\&A, 372, 173

\bibitem[2001]{Bea01b}
  Burrows A., Hubbard W.B., Lunine J.I.,   Liebert J. 2001, Rev. Mod. Phys.,
   73, 719

\bibitem[1997]{Bea97}
  Burrows A., Marley M., Hubbard W.B., et al.
1997, ApJ, 491, 856

\bibitem[1989]{CCM89}
Cardelli J.A., Clayton G.C.,   Mathis J.S. 1989, ApJ, 345, 245

\bibitem[2000]{Chea00}
Chabrier C., Baraffe I., Allard F.,   Hautschild P. 2000, ApJ, 542, 464

\bibitem[1997]{CG97}
  Chiang E.I. , Goldreich P. 1997, ApJ, 490, 368


\bibitem[2001]{Cea01}
  Chiang E.I., Joung M.K., Creech-Eakman M.J., et al.
2001, ApJ, 547, 1077


\bibitem[1998]{Comea98}
  Comer\'on F., Rieke G.H., Claes P., Torra J.,  Laureijs R.J. 1998,
A\&A, 335, 522

\bibitem[2000]{CNK00}
  Comer\'on F., Neuh\"auser R.,  Kaas A.A. 2000, A\&A, 359, 269

\bibitem[1997]{DM97}
D'Antona F.,  Mazzitelli I. 1997, Mem.Soc.Astr.It., 68, 807

\bibitem[2001]{FC01}
Fern\'andez M., Comer\'on F. 2001, A\&A, 380, 264

\bibitem[1996]{Fea96}
  Fukugita M., Ichikawa T., Gunn J.E., et al.
    1996, AJ, 111, 1748

\bibitem[1992]{GY92}
Greene T.P., Young E.T. 1992, ApJ, 395, 516



\bibitem[1992]{L92}
  Landolt A.U. 1992, AJ, 104, 340

\bibitem[2001]{Lea01}
  Leggett S.K., Allard F., Geballe T.R., Hauschildt P.H.,  Schweitzer A.
   2001, ApJ, 548, 908
\bibitem[2002]{Lea02}
Leggett S.K., Golimowski D.A., Fan X. et al. 2002, ApJ, 564, 452

\bibitem[1998]{Linea98}
Lin D.N.C., Laughlin G., Bodenheimer P.,  Rozyczka M. 1998, Science, 281,2025


\bibitem[2001]{Lucea01}
Lucas P.W.,  Roche P.F., Allard F.,  Hauschild P.H. 2001,  MNRAS, 326, 695

\bibitem[2002]{Lodea02}
Lodieu N, Caux E., Monin J.-L.,  Klotz A. 2002, A\&A, 383, L15

\bibitem[1998]{LR98}
Luhman K.L.,  Rieke G.H. 1998, ApJ, 497, 354
\bibitem[1999]{LR99}
Luhman K.L.,  Rieke G.H. 1999, ApJ, 525, 440

\bibitem[2001]{MLAL01}
Muench A.A., Alves J.A., Lada C.J.,  Lada E.A., 2001, ApJ, 558, L51




\bibitem[2001]{Nea01}
  Natta A., Prusti T., Neri R., Wooden D.,  Grinin V.P. 2001, A\&A, 371, 186

\bibitem[2001]{NT01}
Natta A.,  Testi L. 2001, A\&A, 367, L22 (Paper I)


\bibitem[1999]{OTS99}
Oasa Y., Tamura M.,  Sugitani K. 1999, ApJ, 526, 336

\bibitem[2000]{O00}
Oliva E. 2000, Mem. Soc. Astron. Italiana, Vol.~71, p.~861

\bibitem[2000]{PS00}
Palla F.,  Stahler S.W. 2000, ApJ, 540, 255

\bibitem[2001]{PT01}
Papaloizou J.C.B.,  Terquem C. 2001, MNRAS, 325, 221

\bibitem[2000]{Pea00}
  Persi P., Marenzi A.R., Olofsson G., et al. 2000, A\&A, 357, 219


\bibitem[2001]{RC01}
  Reipurth B., Clarke C.J. 2001, AJ, 122, 432

\bibitem[1987]{Shu87}
Shu F.H., Adams F.C.,  Lizano S. 1987, ARA\&A, 25,23

\bibitem[1995]{SKS95}
Strom K.M., Kepner J.,  Strom S.E. 1995, ApJ, 438, 813

\bibitem[2001]{Tea01}
  Testi L., D'Antona F., Ghinassi F., et al. 2001, ApJ, 552, L147

\bibitem[2002a]{Tea02a}
  Testi L., Natta A., Oliva E., et al. 2002a, ApJ, 571, L155 (Paper II)

\bibitem[2002b]{Tea02b}
  Testi L., et al. 2002b, A\&A in preparation


\bibitem[1999]{WGM99}
Wilking B.A., Greene T.P.,  Meyer M.R. 1999, AJ, 117, 469

\end{thebibliography}
\end{document}